\documentclass[doublecol]{epl2}

\usepackage{graphicx}
\usepackage{amsmath}
\usepackage{amssymb}
\usepackage{latexsym}
\usepackage{color}
\usepackage{bm}
\usepackage[font=small,labelfont=bf,
   justification=centerlast ]{caption} 
\usepackage{subcaption}
\usepackage{verbatim}
\usepackage{csquotes}
\usepackage{bibentry}
\usepackage{hyperref}

\renewcommand{\vec}[1]{\bm{#1}}

\DeclareMathOperator{\arctanh}{arctanh}

\graphicspath{{Figures/}}

\title{Equilibrium configurations of hard spheres in a cylindrical harmonic potential}
\shorttitle{Equilibrium configurations of hard spheres in a cylindrical harmonic potential} 

\author{J. Winkelmann\inst{1} \and A. Mughal\inst{2} \and D. Weaire\inst{1} \and S. Hutzler\inst{1}}
\shortauthor{J. Winkelmann \etal} 

\institute{
\inst{1} School of Physics, Trinity College Dublin, The University of Dublin, Ireland \\
\inst{2} Department of Mathematics, Aberystwyth University, Penglais, Aberystwyth, Ceredigion, Wales, SY23 3BZ
}

\abstract{
A line of hard spheres confined by a transverse harmonic potential, with hard walls at its ends, exhibits a variety of buckled structures as it is compressed longitudinally.
Here we show that these may be conveniently observed in a rotating liquid-filled tube (originally introduced by Lee \emph{et al.} [T. Lee, K. Gizynski, and B. Grzybowski, Adv. Mater. {\bf 29}, 1704274 (2017)] to assemble ordered three dimensional structures at higher compressions).
The corresponding theoretical model is transparent and easily investigated numerically, as well as by analytic approximations.
Hence we explore a wide range of predicted structures occurring via bifurcation, of which the stable ones are also observed in our experiments.
Qualitatively similar structures have previously been found in trapped ion systems.
}


\begin{document}
\nobibliography*

\maketitle

\section{Introduction}
Particles confined in the vicinity of a straight line by a transverse potential have been found to exhibit a rich variety of structures, particular distorted linear chains \cite{dessup2015linear}.
In detail they depend on the interactions between the particles, the confining potential and any boundary conditions at the end of a finite sample.
Previous observations have been made with ion traps \cite{mielenz2013trapping,  pyka2013topological, straube2013zigzag, thompson2015ion, partner2015structural, nigmatullin2016formation, yan2016exploring}, but also finite dust clusters \cite{DustCluster2006}, overdamped colloidal systems \cite{Straube_2011} and microfluidic crystals comprising of droplets \cite{Beatus2006}.
The complex scenario of the appearance of such structures, induced by bifurcations upon increasing compression, has been sketched by Landa {\em et al.} \cite{landa2013structure} for the case of quadripolar confining potential and Coulomb interactions.

Here we introduce a much more elementary experimental system for such an investigation.
Our theoretical model is amenable to an analysis using simple numerical and analytical methods.

The experimental system consists of $N$ equal-size hard polypropylene spheres in a horizontal liquid-filled tube, rotating in a lathe.
The spheres are buoyant, so that a centripetal force drives them towards the central axis; the rotational speed is high enough to make gravity negligible.
This set-up was first introduced by Lee \emph{et al.} \cite{Koreans2017} in order to determine the equilibrium phases (ordered three dimensional structures) over a wide range of dimensionless compressions~$\Delta$
\begin{equation}
\Delta = (Nd - L) / d = N - L / d\,,
\label{e:compression}
\end{equation}
where $d$ is the sphere diameter and $L$ is the tube length.
The present application is at very low compression, where the linear chain of contacting spheres is
observed to buckle.
In this range, the structures are all \emph{planar}.
This is assumed in the stepwise method described below, but not in the method of energy minimisation, which indeed finds planar configurations.

The rich scenario of structural transitions within this regime under increasing compression is predicted in detail by the analysis provided below.
It uses two numerical methods: an iterative stepwise solution for force equilibrium positions and a simulation based on energy minimisation.

\section{Theory and numerical analysis}
\subsection{Iterative stepwise method}

\begin{figure*}[h!]
\begin{center}
\includegraphics[width=\textwidth]{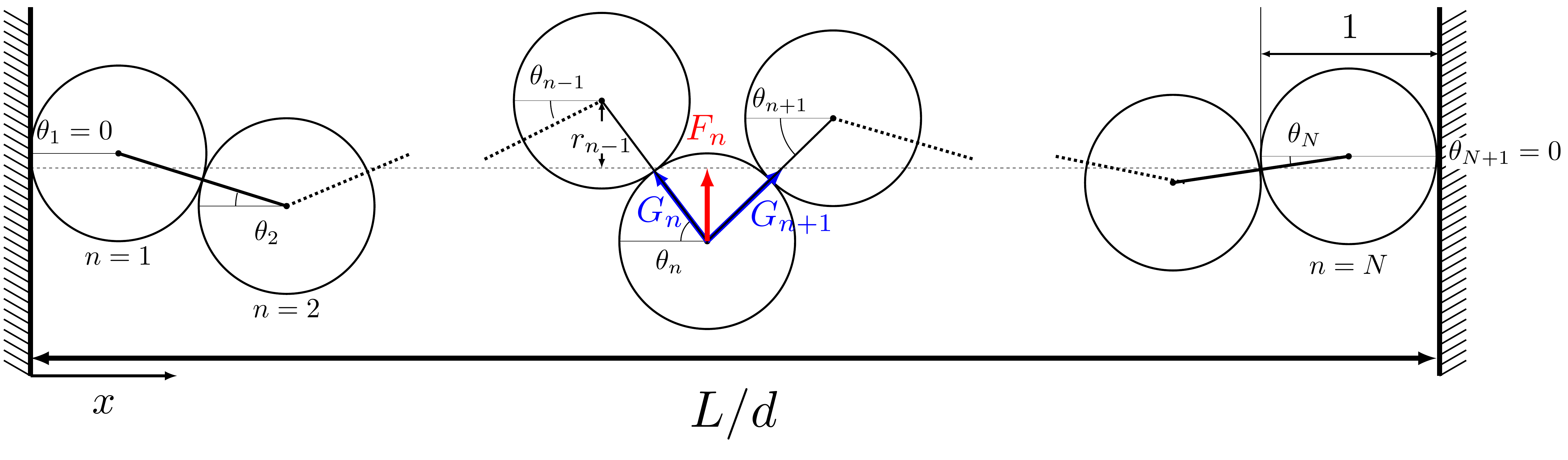}
\caption{
Arrangement of spheres at the two walls and the interior of a \emph{modulated zigzag} structure that is formed when $N$ spheres are compressed between hard walls, showing the notation used for the stepwise solution.
Each sphere, displaced from the central axis (horizontal dashed line) by the dimensionless distance $r_n$, experiences a dimensionless centripetal force $F_n = r_n$ pulling it towards this axis and a compressive force $G_n$.
$\theta_n$ is the angle between the line connecting the centres of spheres $n - 1$ and $n$ and the central axis.
At the wall the line of contact is in the $x$ direction, i.e. $\theta_1= \theta_{N + 1} = 0$. 
}
\label{fig:zigzag}
\end{center}
\end{figure*}

In the experimental system, detailed below, each sphere of mass $m$ experiences a centripetal force $f_c = m \omega^2 R$, where $R$ is the distance of its centre from the central axis of the tube and $\omega$ the rotational speed.
For the case of no compression ($\Delta = 0$) the spheres align in a linear chain along the central axis.
At a finite compression ($\Delta > 0$) the chain starts to buckle and the equilibrium structures take the form of {\em modulated} zig-zag structures as illustrated in Fig \ref{fig:zigzag}.
We developed an elementary stepwise method to describe such structures for low energies. 

In the following we will use the dimensionless distance from the central axis for each sphere $r = R / d$ and the dimensionless centripetal force $F = f_c / (m \omega^2 d) = r$. 

Our aim is to calculate the dimensionless forces $F_n = r_n$ and tilt angles $\theta_n$, as defined in Fig \ref{fig:zigzag}, for $n=1$ to $N$ spheres.
Considerations of force equilibrium and geometrical equations yield iterative relations for $F_n$ (or $r_n$) and $\theta_n$ as follows.

The compressive forces $G_n$ between contacting spheres, are given by $G_n \cos \theta_n = G_0$ from the condition of force equilibrium in the $x$ direction, which is that of the central axis.
$G_0$ is the magnitude of the compressive force at each end of the system.

The equilibrium of centripetal forces $F_n$ on the $n$th sphere gives
\begin{align}
F_n& = G_n \sin\theta_n+G_{n + 1} \sin\theta_{n + 1} \nonumber\\
&=G_0(\tan \theta_{n} +\tan \theta_{n + 1}).
\end{align}
The centres of contacting spheres are separated by their diameter.
Hence the radial distances and forces are
\begin{align}
r_n + r_{n + 1} &= \sin\theta_{n + 1}\,, \nonumber\\
F_n + F_{n + 1} &= \sin\theta_{n + 1}\,.
\end{align}
The above equations then relate $\theta_{n+1}$ and $F_{n+1}$ to $\theta_n$ and $F_n$, i.e.
\begin{align}
\theta_{n+1} &= \arctan\left(\frac{F_n}{G_0} - \tan
\theta_n\right),\nonumber \\
F_{n+1} &= \sin\left[\arctan\left(\frac{F_n}{G_0}-\tan\theta_n\right)\right] - F_n.
\label{e:iterative_nonlin}
\end{align}

These equations may be used in a ``shooting method'' to find solutions for a specified value of $G_0$.
The hard-wall boundary condition for sphere $n=1$ requires the first tilt-angle $\theta_1$ to be zero, with an arbitrary $F_1$.
Using eqs. \eqref{e:iterative_nonlin} we proceed iteratively to $(F_{N + 1}, \theta_{N + 1})$.
The angle $\theta_{N + 1}$ corresponds to the contact of the $N$th sphere with the wall, as illustrated in Fig \ref{fig:zigzag}.

We search for values of $F_1$ (in general more than one) such that the angle $\theta_{N + 1}$ is zero, satisfying the second hard-wall boundary conditions.
This search is performed by coarse graining the initial force $F_1$ over a range of $0 < F_1 \le 0.01$ in steps of $10^{-4}$.
These values are then used as brackets in a bisection method.

The non-dimensional total energy $E$ of such a hard sphere structure can be calculated as
\begin{equation}
E = \frac{E_{\text{rot}}}{m \omega^2 d^2} = \frac{1}{2} \sum_{n=1}^N r_n^2\,,
\label{e:energy} 
\end{equation}
where as in \cite{Winkelmann2019} we have omitted the (constant) energy contribution due to the moment of inertia of the spheres.
The compression $\Delta$ from eq. \eqref{e:compression} is given by
\begin{equation}
\Delta = N - \sum_{n = 1}^N \cos\theta_n\,.
\end{equation}
By performing the root search at varying compressive force $G_0$, we can accumulate a data set, for which we can calculate the energies and compressions in this way.

In practice we encounter difficulties with the stepwise method beyond a compression of $\Delta \ge 0.9$ (for $N=20$).
Above this point our implementation of the bisection search method has problems to find all solutions. 
We expect to successfully extend the application of the stepwise method to this regime in the future.

\subsection{Simulations based on energy minimisation}
\label{sec:simulation}
To confirm and supplement the results of the stepwise method we also seek equilibrium configurations using energy minimisation starting from random configurations.
These simulations are more general than the stepwise method since they are \emph{not} restricted to be planar.
Results for larger compressions can also be generated.
We have used energy minimisation on a system of \emph{soft} spheres, extrapolating to the limit of hard spheres, in order to corroborate the results of the stepwise calculations.

In the soft sphere model, the overlapping spheres repel each other according to Hooke's law with a spring constant $k$.
This crude formalism is often used in the context of foam structures and rheology \cite{Durian95}.

The non-dimensional total energy $E_S$ for $N$ \emph{soft} spheres (of diameter $d$), longitudinally confined between length $L$ is given by, 
\begin{align}
E_S &= \frac{1}{2}\sum_{n=1}^{N}r_n^2 + \frac{1}{2} \left(\frac{k}{m \omega^2}\right)
                                              \left(\sum_{\substack{n, m = 1 \\ m < n}}^{N}\left(\frac{\delta_{nm}}{d}\right)^2 \right. \nonumber\\
                                         &\left. + \left[ \left(\frac{\delta_1}{d}\right)^2 + \left( \frac{\delta_N}{d}\right)^2\right]\right)\,.
\end{align}
The first term is the rotational energy of each sphere.
The second term accounts for the overlap between any two spheres, where the overlap between spheres $n$ and $m$ is defined as $ \delta_{nm} = \vert \vec{R}_n - \vec{R}_m \vert - d$,  where $\vec{R}_n$ and $\vec{R}_m$ are the centre positions of two contacting spheres.
The final term accounts for the overlaps $\delta_1$ and $\delta_N$ of the two end spheres with the two boundaries.

For any given values of compression $\Delta$ and  $k/m \omega^2$ we find equilibrium solutions (stable or metastable) by varying the coordinates of the sphere centres.
Finally, by performing a series of simulations with increasing values of $k/m \omega^2$ we can extrapolate to the hard sphere case (i.e. $ k/m \omega^2 \rightarrow \infty$) and compare directly with the stepwise method. 

The solutions from the stepwise method are only in force equilibrium, i.e. they can be stable or unstable solutions.
We have also used energy minimisation to check the stable/unstable character of the solutions.

\section{Numerical results}
We present results for a variety of low energy structures.
Many other equilibrium structures could be found with higher energy, if we were to extend the range of our search parameters.
These may be of limited physical significance and will not be pursued in the present paper.

\subsection{Typical profiles}

\begin{figure}[h!]
\centering
\includegraphics[width=\linewidth]{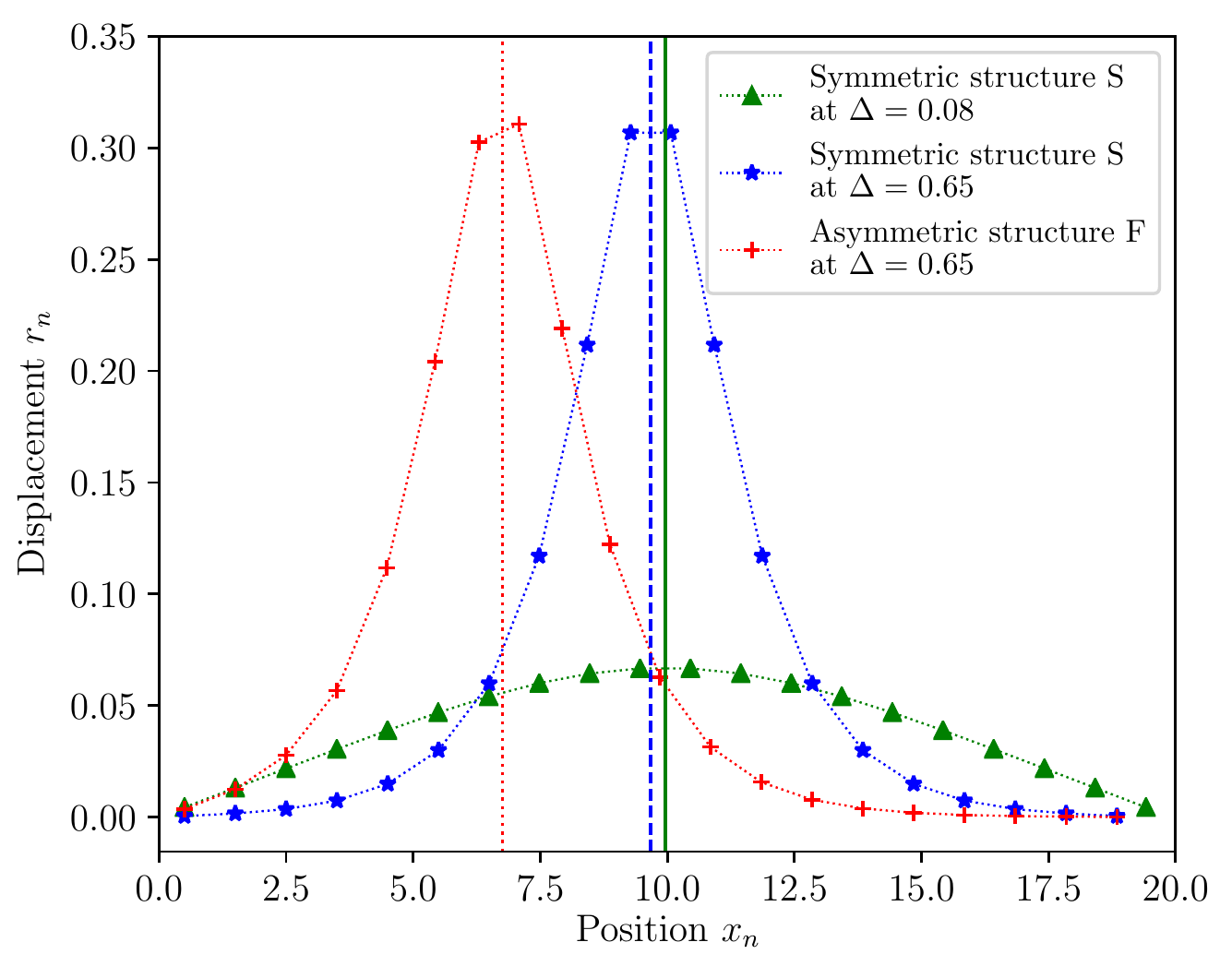}
\caption{
Sphere displacements $r_n$ as function of position $x_n$ for a symmetric (blue stars) and a stable asymmetric structure (red crosses) at a compression of $\Delta = 0.65$.
Also shown is the displacement $r_n$ for the symmetric structure at a lower compression of $\Delta = 0.08$ (green triangles).
(All quantities are dimensionless, see definitions in main text.)
The peak position of the asymmetric structure for $\Delta = 0.65$, estimated by a quadratic fit of the displacements around the maximum, is displayed by the vertical dotted red line.
The vertical blue dashed and the green solid line display the midpoint of the system.
The distance between $x_0$ and $x_N$ is equal to $N - \Delta - 1$. 
}
\label{fig:adilcomparison}
\end{figure}

For low compressions our search yields only one structure that we will refer to as the \emph{symmetric} structure S, since the profile for $F_n$ (or displacement $r_n$) is symmetric around the midpoint of the system. 
(Note that we have defined $F_n$ and $r_n$ to be positive.)

Examples of such a profile for $N = 20$ are presented in Fig \ref{fig:adilcomparison} for a low (green triangles) and high (blue stars) compression where we show the displacement $r_n$ from the central axis vs the (dimensionless) position $x_n~=~1/2~+~\sum_{i = 2}^n~\cos(\theta_i)$.

These results show perfect agreement with the symmetric structure generated by energy minimisation and extrapolated to the hard sphere limit.
The structures obtained by energy minimisation are necessarily confined to stable cases.

For high compressions additional asymmetric structures are obtained from the stepwise method.
An example for the displacement profile for such a structure is given by the red crosses in Fig \ref{fig:adilcomparison}.

\subsection{Bifurcation diagrams}

\begin{figure}[h!]
\centering
\includegraphics[width=0.95\linewidth]{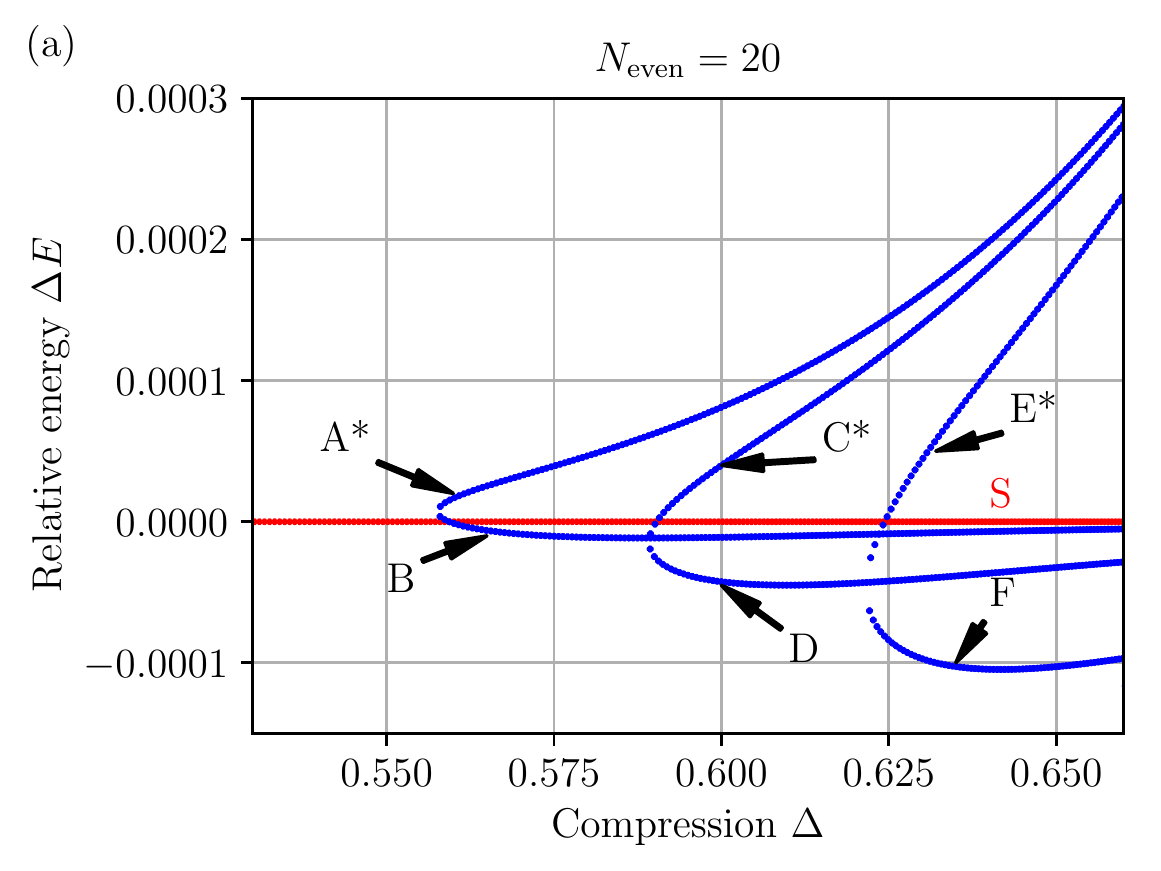}
\includegraphics[width=0.95\linewidth]{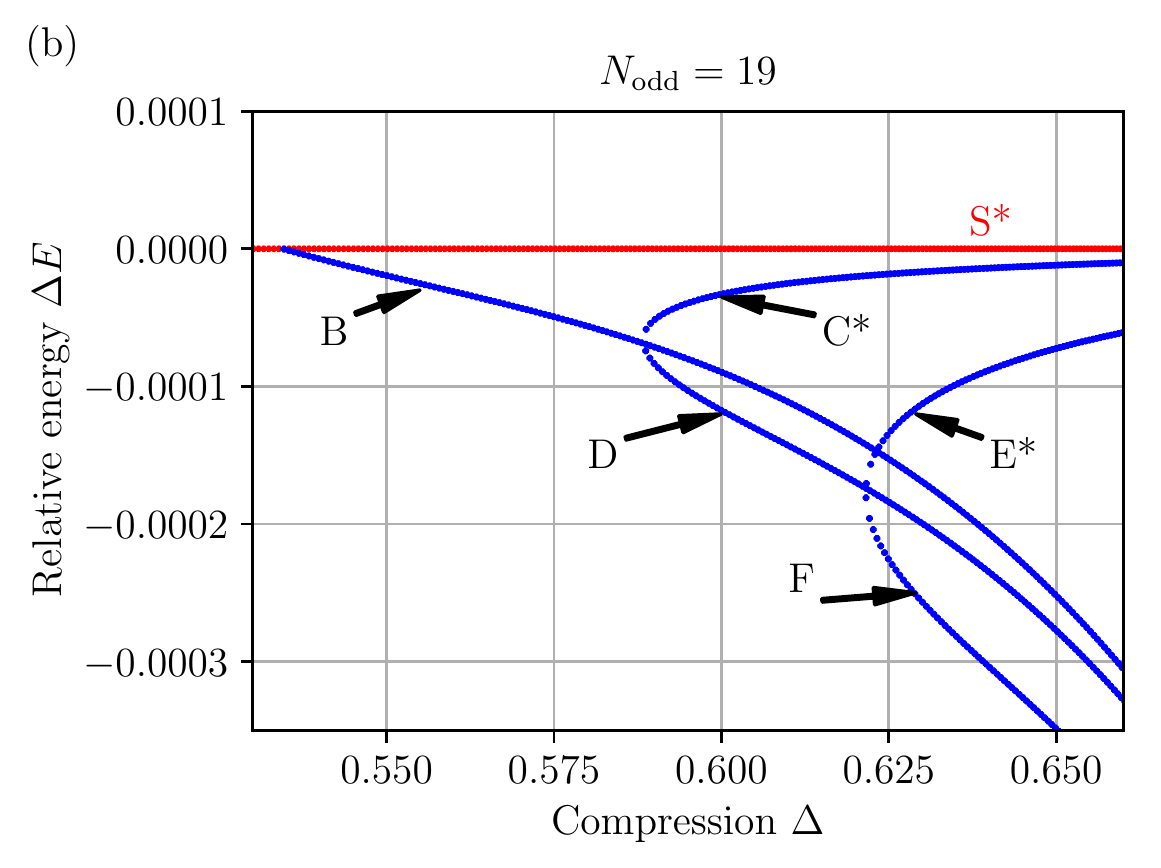}
\caption{
Bifurcation diagram: Relative energies $\Delta E = E - E_{\text{Symm}}$ (where $E_{\text{Symm}}$ is the energy of the symmetric structure S), are plotted against compression $\Delta$ around the first bifurcation for the case of even number of spheres (a) and odd number of spheres (b).
Unstable structures are marked with an asterisk.
Examples of all structures in the even case are displayed in Fig~\ref{fig:theorystructures}.
}
\label{fig:energies}
\end{figure}

We have used the iterative stepwise method to search for low energy structures in the range of the compressive forces between $0.2 \le G_0 \le 0.25$ and initial forces between $0 < F_1 < 0.01$.
These structures correspond to relative compression below $\Delta < 0.9$.
They were computed for both an even ($N = 20$) and an odd ($N = 19$) number of spheres, for which the results differ qualitatively.

In these parameter ranges the total energy $E_{\text{Symm}}$ of the symmetric structure S increases from $0$ to roughly $0.15$, whereas the difference between the energies of alternative structures is only of order $10^{-4}$.
We therefore computed the energy $\Delta E = E - E_{\text{Symm}}$ relative to that of the symmetric structure at the same compressive force $G_0$ and plotted them against their compression for even (Fig \ref{fig:energies}(a)) and odd case (Fig \ref{fig:energies}(b)).

We present these relative energies in the vicinity of the compression range where the first asymmetric structures are created by bifurcation.
While both cases of even and odd $N$ feature an increasing number of bifurcations as compression is increased, they are qualitatively different and will thus be discussed separately.

For the even case $N = 20$ an increasing number of asymmetric structures (A-F) are introduced by bifurcation as compression increases.
The unstable structures are marked with an asterisk.
The first two additional branches A* and B emerge from an \enquote{out-of-the-blue} bifurcation at $\Delta = 0.558$ without any preceding structure.
Of these two branches, structures on branch B are stable, whereas structures from A* are unstable, as verified by energy minimisation.

Two further structures C* and D, appear via a pitch-fork bifurcation out of the previous stable structure B at $\Delta = 0.588$.
B and the additional branch of lower energy D are stable, whereas the upper one C* is not.
A similar pitch-fork bifurcation of the D branch occurs for the next two structures at $\Delta = 0.622$, from which the lower branch F is again stable and E* unstable.

\begin{figure}[h!]
\centering
\includegraphics[width=\linewidth]{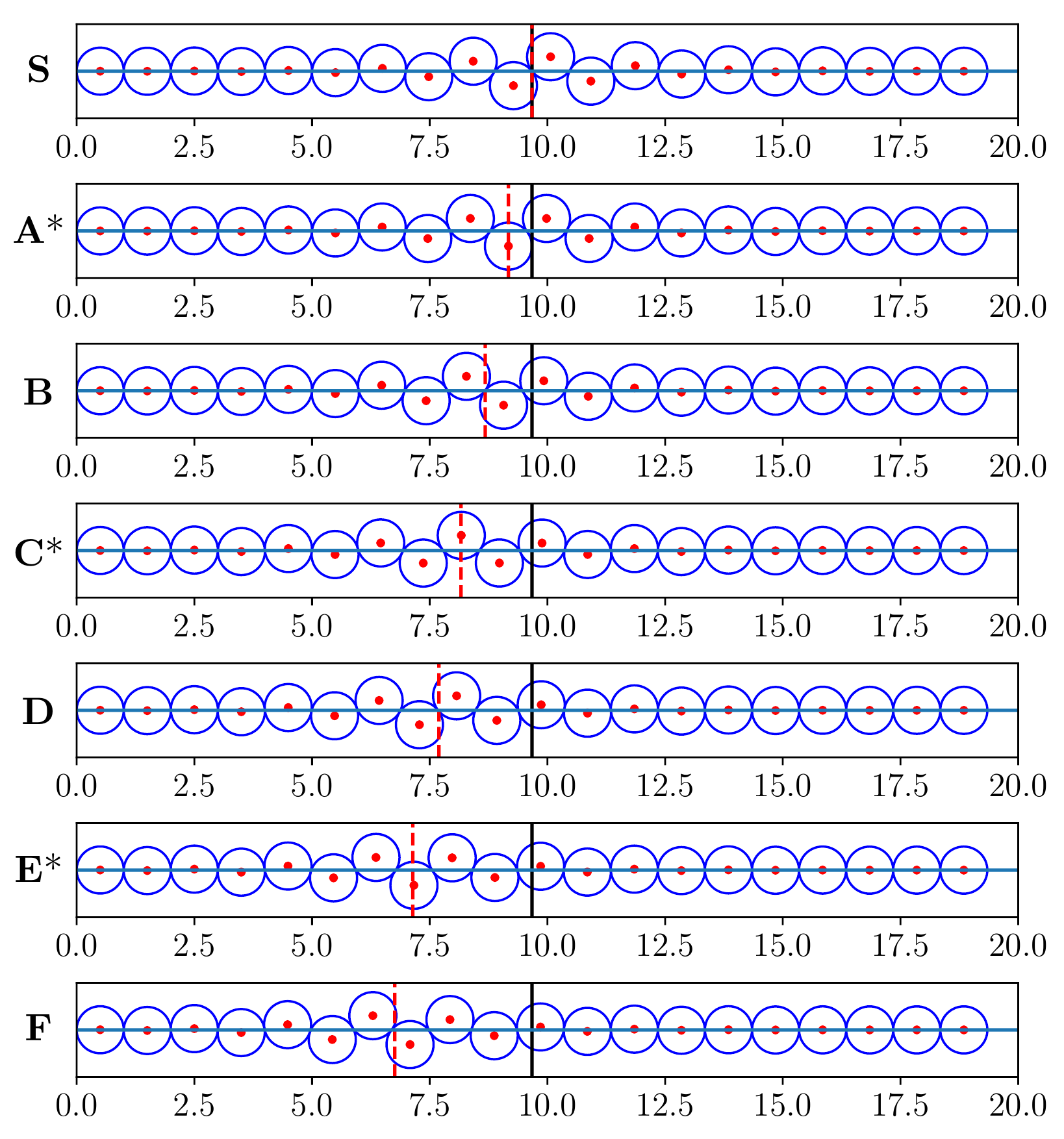}
\caption{
Examples of buckled chain structures from the S and A-F branches, as labelled in Fig \ref{fig:energies}.
The structures were created by the stepwise method with $N = 20$ spheres at a compression of $\Delta = 0.65$.
Structures marked by an asterisk are unstable.
The solid black vertical line marks the midpoint of the system, while the dashed red vertical line marks the peak position of the position profile.
Asymmetric structures (A-F) are doubly degenerate (i.e. can have a peak on the left or on the right of the centre).
}
\label{fig:theorystructures}
\end{figure}

Examples of structures from all of the seven branches for the even case in Fig \ref{fig:energies}(a) are given in Fig \ref{fig:theorystructures}.
The vertical black solid line in these plots represents the midpoint of the structure;
the vertical red dashed line indicates the peak position of the sphere profile, as estimated by a quadratic fit to the sphere positions around the maximum.
For unstable structures the peak position coincides roughly with the centre position of a sphere.
Note the degeneracy: asymmetric structures may have the peak left or right of the centre.

The energy diagram for the odd case of 19 spheres (in Fig \ref{fig:energies}(b)) differs with respect to the first bifurcation.
Here only a single new stable structure (branch B) emerges.
From then on bifurcations follow the pattern of the even case, in which previous structures remain stable and new structures of lower energy are stable (i.e. D and F are stable, while C* and E* are unstable).

While the structures that we have identified here appear to be the only equilibria within the specified range in energy and compression, structures with a more complicated profile occur at higher energy, which we have not addressed here.
The displacement profiles of these structures can contain two or more off-centred peaks.

\subsection{Maximum angles}

We have also computed the maximum angle $\theta_{\text{max}}$ of the symmetric structure with varying compression for the stepwise method and energy minimisation, see Fig \ref{fig:thetas}.
This is a quantity that can readily be extracted from experimental data, see below.

While our results for the stepwise method stop at a compression of $0.9$, the maximum angle $\theta_{\text{max}}$ from the energy minimisation was computed up to a compression of $\Delta \lesssim 1.3$.
At this point the modulated zigzag structure acquires an additional contact with the next-nearest neighbour sphere.

At low compressions ($\Delta \lessapprox 0.1$), where the displacement profile is of the type shown by the green triangles in Fig \ref{fig:adilcomparison},  $\theta_{\text{max}}$ varies as $\sqrt{\Delta}$.

\begin{figure}[h!]
\centering
\includegraphics[width=\linewidth]{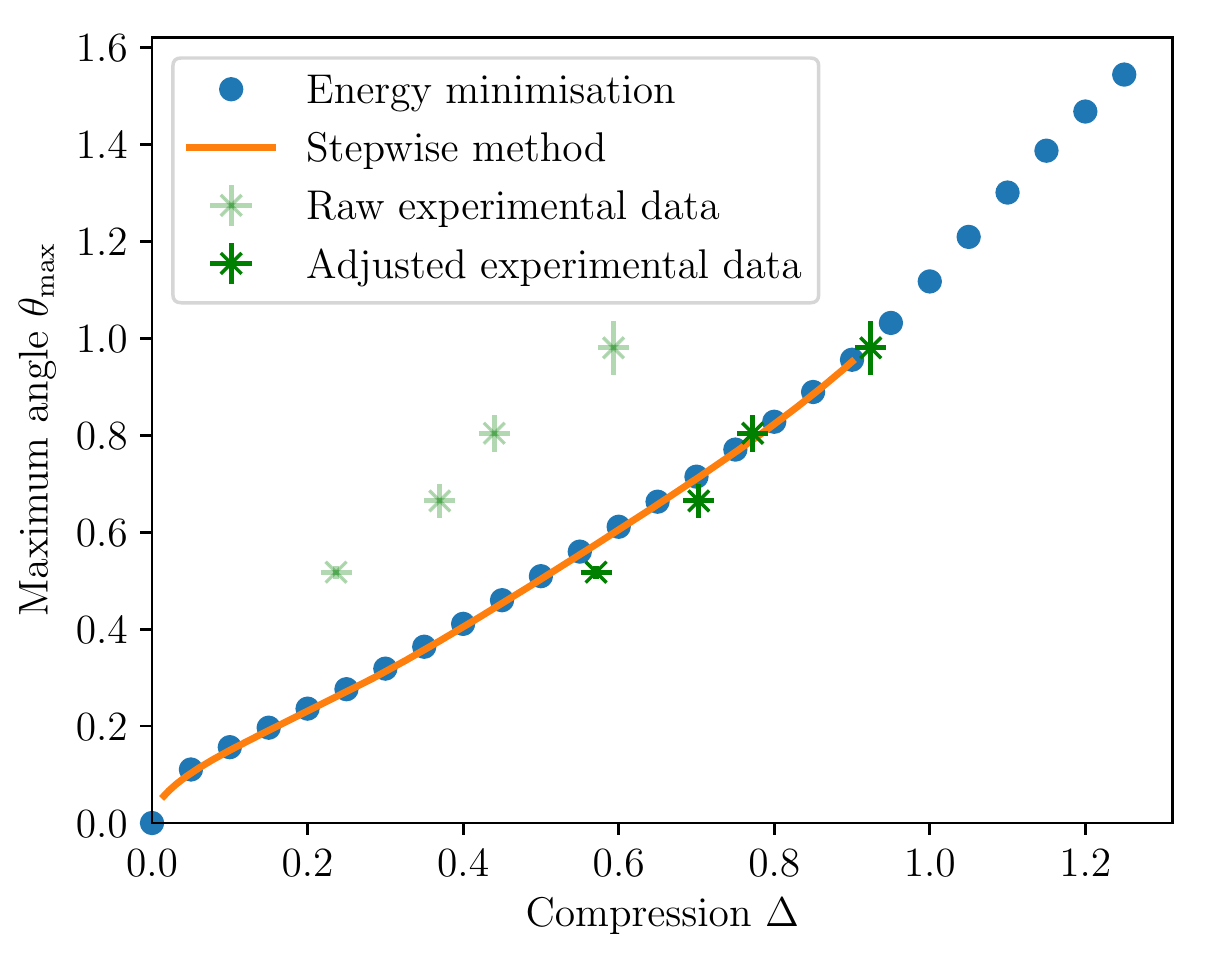}
\caption{
Maximum angle $\theta_{\text{max}}$ of the symmetric structure as a function of compression $\Delta$ for an even number of spheres $N$.
Blue circles and the orange solid line are obtained from numerical calculations (stepwise method and energy minimisation).
The green crosses with low opacity refer to the raw experimental data points.
For the green crosses with high opacity, the increased effective sphere diameter attributed to vibrations in the system was taken into account in the compression calculation.
The uncertainty in the value for $\theta_{\text{max}}$ was obtained by averaging the angles over five images of the structure at the same compression.
}
\label{fig:thetas}
\end{figure}

\section{Linear approximation}
\label{sec:linear}

\begin{figure}
\centering
\includegraphics[width=\linewidth]{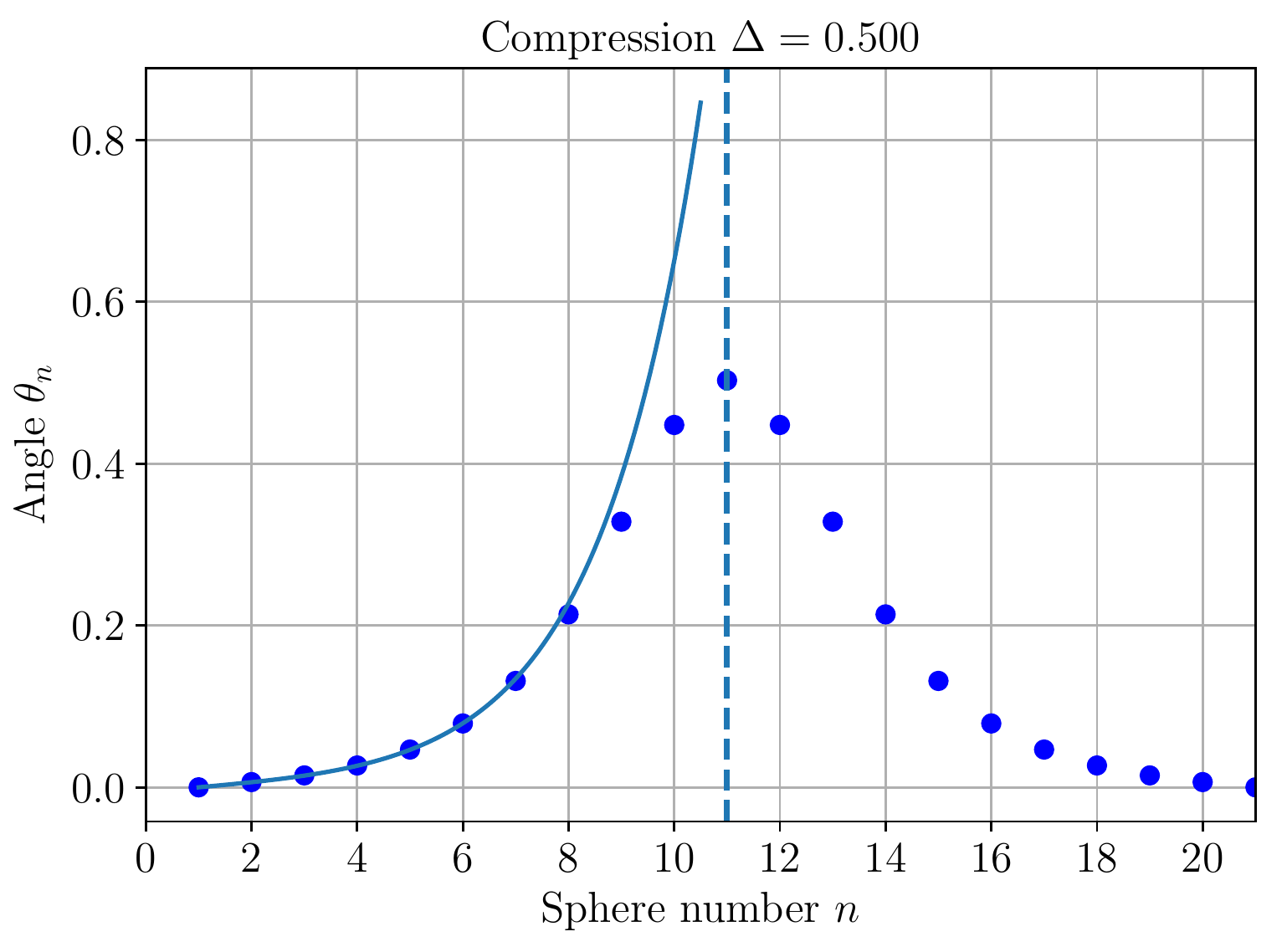}
\caption{
Variation of angle $\theta_n$ with sphere number $n$ as obtained from the exact stepwise method (blue dots) and from the linearised theory \eqref{e:theta} (blue solid line) (for $N = 20$) at a compression of $\Delta = 0.500$.
Note that there are $21$ angles, since $\theta_{N + 1}$ is associated with the wall contact of sphere $N$.
}
\label{fig:linearised}
\end{figure}

In order to better understand the above results, we have developed an approximate, linear analytic description as follows.
For small angles $\theta_n$ and forces $F_n$, linearisation of eq. \eqref{e:iterative_nonlin} leads to
\begin{equation}
\begin{pmatrix}
    F_n\\
    \theta_n
  \end{pmatrix}
  =
  \begin{pmatrix}
    \frac{1}{G_0} -1& -1\\
    \frac{1}{G_0} & -1
  \end{pmatrix}
  \begin{pmatrix}
    F_{n-1}\\
    \theta_{n-1}
  \end{pmatrix}.
\end{equation}

Recursive substitution of $F_n$ and $\theta_n$ and setting $\frac{1}{G_0}~=~4~+~\epsilon$ for $\epsilon$ small and positive, results in:
\begin{equation}
\begin{pmatrix}
    F_n\\
    \theta_n
  \end{pmatrix}
  =
  \begin{pmatrix}
     3 + \epsilon& -1\\
    4 + \epsilon & -1
  \end{pmatrix}^{n - 1}
  \begin{pmatrix}
    F_{1}\\
    \theta_{1}
  \end{pmatrix}.
\end{equation}
The largest possible value for the compressive force is $G_0 = 1/4$, which, for the case of an infinitely long chain, corresponds to the uniform zigzag structure ($r_n = -r_{n-1}$).

A solution for $(F_n, \theta_n)^T$ may be expressed in terms of eigenvalues $\lambda_{1, 2}$ and eigenvectors $\vec{V}_{1, 2}$ of the above matrix as $(F_n, \theta_n)^T = a \lambda_1^{n - 1} \vec{V}_1 + b \lambda_2^{n - 1} \vec{V}_2$.
To lowest order in $\epsilon$ the eigenvalues are given by $\lambda_{1,2}=1 \pm \sqrt{\epsilon}$ with the corresponding eigenvectors $\vec{V}_{1,2}= \left(\frac{1}{2}(1\pm \frac{\sqrt{\epsilon}}{2}), 1\right)^T$.
The prefactors $a$ and $b$ are obtained from the initial conditions $F_1$ and $\theta_1$.

The solution in the linearised approximation for $\theta_1 = 0$ is then given by
\begin{align}
F_n &= \frac{F_1}{\sinh(\phi)} \sinh(\sqrt{\epsilon} (n - 1) + \phi) \\
\theta_n &= \frac{4 F_1}{\sqrt{\epsilon}} \sinh(\sqrt{\epsilon} (n - 1))\,,
\label{e:theta}
\end{align}
with the offset in the forces $\phi~=~\arctanh(\sqrt{\epsilon} / 2)$.
Note that $\theta_n$ does not have an offset, since $\theta_1 = 0$.

A comparison of angles $\theta_n$ using the approximated linearised equation \eqref{e:theta} and the previously numerical exact stepwise method is shown in Fig \ref{fig:linearised} for a compressive force of $G_0 = 0.234$, resulting in a compression of $\Delta = 0.500$.
The starting value $F_1$ in the linearised scheme was taken from the corresponding value in the stepwise method.

We find excellent agreement between the linearised theory and the stepwise method up to about $n = 8$.
The linear theory produces a monotonically increasing function (Fig \ref{fig:linearised}), whereas the accurate solution \enquote{rolls over} and decreases towards the second boundary.
This can be understood in terms of the role of nonlinearity, and approximated in an \emph{ad hoc} manner: we will leave this to a subsequent paper.

\section{Comparison with experiment}
\label{sec:experiments}

Our experimental procedure is similar to that of Lee \textit{et al.} \cite{Koreans2017}.
We placed an even number of $N = 34$ polypropylene beads of density $\rho = 0.900\,\text{g/cm$^3$}$ and diameter $d = 3.000 \pm 0.001\,\text{mm}$  \cite{manufacturer} in a cylindrical tube (inner diameter $15.91 \pm 0.01\,\text{mm}$; outer diameter $20.17 \pm 0.01\,\text{mm}$; length $130.55 \pm 0.01\,\text{mm}$) filled with water (density $\rho_w = 1\,\text{g/cm$^3$}$).

\begin{figure}[h!]
\centering
\includegraphics[width=\linewidth]{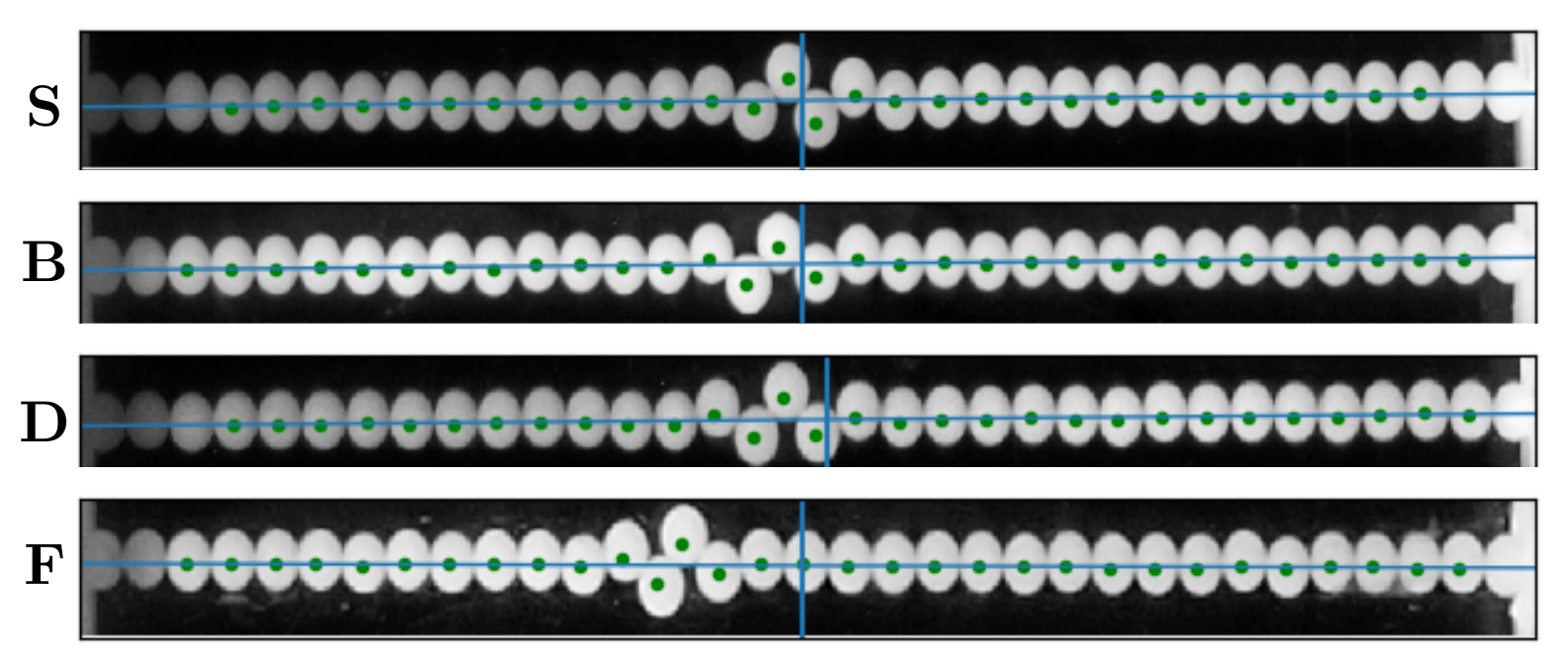}
\caption{
Photographs of a variety of buckled structures for $N = 34$, obtained by the rapid rotation of a water-filled tube containing polypropylene beads (density $\rho = 0.900 \,\text{g/cm$^3$}$), using a lathe.
The structures S, B, D, and F are labelled as in Fig \ref{fig:energies}.
The vertical line represents the midpoint of the system and the horizontal line the central axis of rotation.
}
\label{fig:experiments}
\end{figure}

The tube is sealed on both ends with stoppers, making sure that no air bubbles remain within the system.
The extent to which the stoppers intrude into the tube can be varied, allowing us to adjust the compression $\Delta$.

The tube is then mounted onto a commercial lathe (Charnwood W824), for which we set the rotation frequency to $\omega = 1800 \pm 50\,\text{rpm}$.
In order to record the structures we used a stroboscopic lamp, whose frequency is matched to that of the lathe.
A slight off-set between both frequencies is used so that recorded structures appear to be slowly rotating (see example in the supplemental video \href{https://www.maths.tcd.ie/~jwinkelm/Files/EPLSupplementalMaterial/LatheExperiments.mp4}{LatheExperiment.mp4}).

Fig \ref{fig:experiments} shows images of the structures that we identified with the branches S, B, D, and F by comparing them with to the numerical results from \ref{fig:theorystructures}.
The experimental structures can be identified by the distance of the peak position to the structure centre.
This is independent of $N$ for large number of spheres because the wall effects can be neglected. 
The identified structures correspond to all the stable structures of Fig \ref{fig:energies}.
Structure S, as well as structure B, were found at a compression of $\Delta = 0.44 \pm 0.02$, while the compressions for structure D was $\Delta = 0.59\pm0.02$ and for F, $\Delta = 0.68 \pm 0.02$.

However, in order to reconcile these experimental results with the theoretical predictions of previous sections, it is necessary to introduce an {\em effective} diameter for the spheres, about $1\,\%$ greater than the true value.
This increases the effective compressions by a constant shift of roughly $0.35$. 
We attribute it to the effects of vibration of the lathe, and will explore strategies for its mitigation in future work.
This shift also features in previous results from Lee \textit{et al.} \cite{Koreans2017, Winkelmann2019}.

We extracted the maximum angle $\theta_{\text{max}}$ for the symmetric structure with varying compression for the experiments (see Fig \ref{fig:thetas}).
It shows very clearly the necessity for the adjustment of sphere diameter.
Due to the neglecting wall effects, these results only depend on $N$ being odd or even for a large enough number of spheres.

\section{Conclusion}
The compressed and confined sphere chain presents a variety of fascinating observations, previously described in terms of ``kinks'' or ``solitons'' \cite{partner2015structural}.
We have succeeded in exploring many of its properties, using simple apparatus and theoretical methods.

Recently we have found a yet simpler experimental method which should be useful, at least for purposes of demonstration.
It consists of a horizontal tube into which ball bearings are introduced.
Slight agitation enables them to settle in modulated zigzag structures similar to those depicted above \cite{demonstration}.
A further variation, which appears to be promising, uses bubbles in a liquid-filled tube.

While we have so far investigated only simple structures with single peaks in the displacement profile,
more complicated structures exist at higher energies, and may also be found with the stepwise method.
Among these are structures that can be created by concatenating one of the single-peak structures with its mirrored counterpart.
Their compression and energy will be doubled.

Other extensions will include the case of soft (elastic) spheres, for which we have already observed similar effects, using hydrogel particles and bubbles.
Also the observations can be extended to much higher compression, making contact with the work of Lee \textit{et al.} \cite{Koreans2017} and Winkelmann \emph{et al.} \cite{Winkelmann2019}, for the 3d structures generated.
It may also be possible to take advantage of a technique that uses photoelastic material to indicate the magnitude of the compressive forces \cite{MajmudarPhotoelastic, KarenPhotoelastic}.

Considerable current interest in the compressed sphere chain focusses on motion of kinks and the corresponding Peierls--Nabarro potential.
This may be estimated by making a smooth interpolation of the energy values for stable and unstable states as calculated here.

We hope that the results presented here will find direct comparison with previous work, particularly with regards to ions confined in traps \cite{mielenz2013trapping,  pyka2013topological, landa2013structure,straube2013zigzag, thompson2015ion, dessup2015linear, partner2015structural, nigmatullin2016formation, yan2016exploring} as well as assemblies of magnetic particles in a channel \cite{galvan2014magnetic}. This work should also be relevant to other systems in which buckling is a key feature: for example localised buckling has recently been observed in experiments involving an expanding (growing) elastic beam pinned to a substrate \cite{michaels2019geometric}.

\acknowledgements
We wish to thank S. Burke for help with the experimental set-up.
This work was supported by: EPSRC grant numbers EP/K032208/1 and
EP/R014604/1, as well as an Irish Research Council Postgraduate Scholarship
(project ID GOIPG/2015/1998). We also acknowledge the support of
Science Foundation Ireland (SFI) under grant number 13/IA/1926.


\end{document}